\documentclass[12pt]{article}
\usepackage{amsmath,amsthm,amsfonts,graphicx,epsfig}

\newcommand{\e}{\varepsilon}

\newcommand{\dbar}{\kern-.1em{\raise.8ex\hbox{ -}}\kern-.6em{d}}
 \font\tenrm=cmr10 
\begin{document}
\title{Pushmepullyou: An efficient micro-swimmer}

\author{ J.E. Avron, O. Kenneth and D.H. Oaknin\\
\tenrm\!  Department of
Physics, Technion, Haifa 32000, Israel\\
}%
\maketitle
\begin{abstract} The swimming of a pair of spherical bladders that
change their  volumes and mutual distance is  efficient at low
Reynolds numbers and is superior to other models of artificial
swimmers. The change of shape resembles the wriggling motion known
as {\it metaboly} of certain protozoa.
\end{abstract}

\begin{figure}[htb]
\hskip 2
cm\includegraphics[width=5cm]{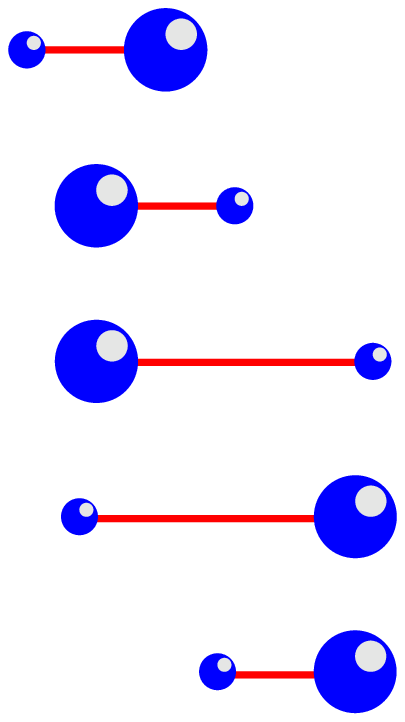}\includegraphics[width=5cm]{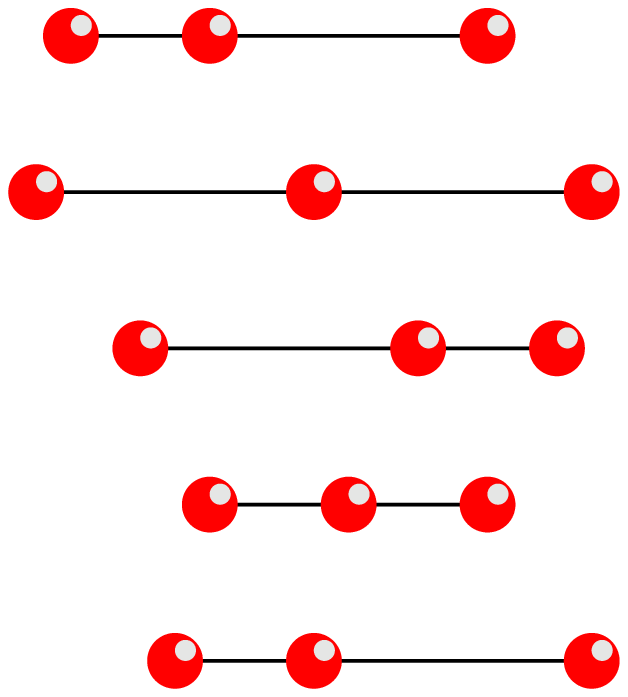}
\caption{\em Five snapshots of the pushmepullyou swimming stroke
(left) and the corresponding strokes of the three linked spheres
(right). Both figures are schematic. After a full cycle the
swimmers resume their original shape but are displaced to the
right. Pushmepullyou is both more intuitive and more efficient
than the three linked spheres.}\label{fig:swimmer}
\end{figure}

Swimming at low Reynolds numbers can be remote from common
intuition because of the absence of inertia \cite{childress}. In
fact, even the direction of swimming may be hard to foretell
\cite{purcel}. At the same time, and not unrelated to this, it
does not require elaborate designs: Any stroke that is not
self-retracing will, generically, lead to some swimming
\cite{wilczek}. A  simple model that illustrates these features is
the  three linked spheres \cite{najafi}, Fig.~\ref{fig:swimmer}
(right), that swim by manipulating the distances $\ell_{1,2}$
between neighboring spheres. The swimming stroke is a closed, area
enclosing, path in the $\ell_1-\ell_2$ plane.  Another mechanical
model that has actually been built is Purcell's two hinge model
\cite{blades}.

Swimming efficiently is an issue for artificial micro-swimmers
\cite{agk}. As we have been cautioned by Purcell not to trust
common intuition at low Reynolds numbers \cite{purcel}, one may
worry that efficient swimming may involve unusual and nonintuitive
swimming styles. The aim of this letter is to give an example of
an elementary and fairly intuitive swimmer that is also remarkably
efficient provided it is allowed to make large strokes.

The swimmer is made of two spherical bladders, Fig.~\ref{fig:swimmer}  
(left). The bladders are elastic bodies which impose no-slip boundary 
conditions. The device swims by cyclically changing the distance between 
the bladders and their relative volumes. For the sake of simplicity and 
concreteness we assume that their total volume, $v_0$, is conserved. The 
swimming stroke is a closed path in the $ v-\ell$ plane where $v$ is the
volume of, say, the left sphere and $\ell$ the distance between them. 
We shall make the further simplifying assumption that the viscosity of the 
fluid  contained in the bladders is negligible compared with the viscosity 
of the ambient fluid.
For reasons that shall become clear below we call the swimmer 
pushmepullyou.


Like the three linked spheres, pushmepullyou is mathematically
elementary only in the limit that the distance between the spheres
is large, i.e. when $\e_i=a_i/\ell\ll 1$. ($a_i$ stands for the
radii of the two spheres and $\ell$ for the distances between the
spheres.) We assume that the Reynolds number $R={\rho av/\mu}\ll
1$, and that the distance $\ell$ is not  too large: ${\ell v}\ll
\mu/\rho$. The second assumption is not essential and is made for
simplicity only. (To treat large $\ell$ one needs to replace the
Stokes solution, Eq.~(\ref{stokes}), by the more complicated, but
still elementary, Oseen-Lamb solution \cite{batchelor}.)


Pushmepullyou is simpler than the three linked spheres: It
involves two spheres rather than three; it is more intuitive and
is easier to solve mathematically. It also swims a larger distance
per stroke and is considerably more efficient \cite{movie}. If
large strokes are allowed, it can even outperform conventional
models of biological swimmers that swim by beating a flagellum
\cite{lighthill}. If only small strokes are allowed then
pushmepullyou, like all squirmers \cite{agk}, becomes rather
inefficient.

The swimming velocity is defined by $\dot X= (U_1+U_2)/2$ where
$U_i$ are the velocities of the centers of the two spheres. To
solve a swimming problem one needs to find the (linear) relation
between the (differential) displacement $\dbar X$, and the
(differential) controls $(d\ell,dv)$. This relation, as we shall
show, takes the form:
\begin{equation}\label{swim}
2\, \dbar X=  \frac {a_1-a_2}{a_1+a_2}\ d\ell\ + \frac 1
{2\pi\ell^2} \ d v,
\end{equation}
where $a_1,a_2$ are the radii of the left and right spheres
respectively and $v$ is the volume of the left bladder. $\dbar X$
stresses that the differential displacement does not integrate to
a function $X(\ell,v)$. Rather, the displacement $X(\gamma)$
depends on the stroke $\gamma$, defined as a closed path in $\ell-v$ 
plane. The first term says that increasing $\ell$ leads to swimming in
the direction of the small sphere. It can be interpreted
physically as the statement that the larger sphere acts as an
anchor while the smaller sphere does most of the motion when the
``piston'' $\ell$ is extended. The second term says that when
$\ell$ is held fixed, the swimming is in the direction of the
contracting sphere: The expanding sphere acts as a source pushing
away the shrinking sphere which acts as a sink to pull the
expanding sphere. This is why the swimmer is dubbed pushmepullyou.

To gain further insight consider the special case of small strokes
near equal bi-spheres. Using Eq.~(\ref{swim}) one finds, dropping
sub-leading terms in $\e_i=a_i/\ell$:
\begin{equation}\label{curvature}
\delta X= \,\frac{1}{6}\, d\log v\wedge d\ell
\end{equation}
The distance covered in one stroke scales like the area in $\log
v-\ell$ plane. Note that the swimming distance {\em does not}
scale to zero with $\e$, when the spheres are far apart. This is
in contrast with the three linked spheres where the swimming
distance of one stroke is proportional to $\e$. For a small cycle
in the $\ell_1-\ell_2$ plane Najafi et. al. find for a symmetric
swimmer (Eq.~(11) in \cite{najafi}):
\begin{equation}\label{curvature-iran}
\delta X=0.7  \e \,d\log \ell_2\wedge d\ell_1
\end{equation}
When the swimmer is elementary, (=when $\e$ is small), it is also
poor.

Consider now a large stroke associated with the closed rectangular
path enclosing the box $\ell_s\leq\ell\leq\ell_L,\; v_s\leq v_1,
v_2 \leq v_L\equiv v_0-v_s$, where $v_1 = v$ and $v_2$ are,
respectively, the volumes of the left and right bladders.  If
$a_s\ll a_L$ then from Eq.~(\ref{swim}), $X(\gamma)$ is
essentially  $\ell_L-\ell_s$:
\begin{equation}\label{step}
X(\gamma) = \left( \frac {a_L-a_s}{a_L+a_s}\right)\ (\ell_L-
\ell_s) \left(1+O(\e^3)\right)
\end{equation}
This says that the distance covered in one stroke is of the order
of the size of the swimmer, i.e. the distance between the balls
$\ell$.

Certain protozoa and species of {\it Euglena} perform a wriggling
motion known as {\it metaboly} where, like pushmepullyou, body
fluids are transferred from a large spheroid to a small spheroid
\cite{metaboly}. Metaboly is, at present not well understood and
while some suggest that it plays a role in feeding others argue
that it is relevant to locomotion \cite{theriot}. The
pushmepullyou model shows that at least as far as fluid dynamics
is concerned, metaboly is a viable method of locomotion. Racing
tests made by R. Triemer \cite{triemer} show that Euglenoids swim
1-1.5 their body length per stroke, in agreement with
Eq.~(\ref{step}) for reasonable choices of stroke parameters.
Since Euglena resemble deformed pears
--- for which there is no known solution to the flow equations ---
Pushmepullyou is, at best, a biological over-simplification. It
has the virtue that it admits complete analysis.

The second step in solving a swimming problem is to compute the
power $P$ needed to propel the swimmer. By general principles, $P$
is a quadratic form in the velocities in the control space and is
proportional to the (ambient) viscosity $\mu$. The problem is to
find this quadratic form explicitly. If the viscosity of the fluid
inside the bladders is negligible, one finds that in order to
drive the controls $\ell$ and $v$, Pushmepullyou needs to invest
the power
\begin{equation}\label{metric}
\frac P {6\pi\mu}=  \left(\frac 1 {a_1} +\frac 1 {a_2}\right)^{-1}
\, \dot\ell^2 +\frac {2} {9\pi}\  \left(\frac 1 {v_1} +\frac 1
{v_2}\right) \dot{v}^2\end{equation} Note that the dissipation
associated with  $\dot\ell$, is dictated by the {\em small} sphere
and decreases as the radius of the small sphere shrinks. ( The
radius can not get arbitrarily small and must remain much larger
than the atomic scale for Stokes equations to hold.) The moral of
this is that pushing the small sphere is frugal. The dissipation
associated with $\dot v$ is also dictated by the small sphere.
However, in this case, dilating a small sphere is expensive.

The drag coefficient is a natural measure to  compare different
swimmers. It measures the energy dissipated in swimming a fixed
distance at fixed speed. (One can always decrease the dissipation
by swimming more slowly.)  Let $\tau$ denote the stroke period.
The drag is formally defined by \cite{lighthill,samuel}:
\begin{equation}\label{delta}
\delta(\gamma)=\frac { \tau\int_0^\tau P dt}{6\pi\mu
X^2(\gamma)}\,.
\end{equation}
$X(\gamma)$ is the swimming distance of the stroke $\gamma$. The
smaller $\delta$ the more efficient the swimmer. $\delta$ has the
dimension of length (in three dimensions) and is normalized so
that dragging of a sphere of radius $a$ with an external force has
$\delta=a$.


To compute the dissipation for the rectangular path we need to
choose rates for traversing it.  The optimal rates are constant on
each leg provided the coordinates are chosen as
$(\ell,\arcsin\sqrt {v\over v_0})$.  This can be seen from the
fact that if we define $x=\arcsin\sqrt{v\over v_0}$,  then
$4v_0\dot{x}^2= \left(\frac 1 {v_1} +\frac 1 {v_2}\right)
\dot{v}^2$ and the Lagrangian associated with Eq.~(\ref{metric})
is quadratic in $(\dot \ell,\dot x)$ with constant coefficients,
like the ordinary kinetic Lagrangian of non relativistic
mechanics. It is a common fact that the optimal path of such a
Lagrangian has constant speed.

>From Eq.~(\ref{metric}) we find, provided also
$\ell_L^2\gg\ell_s^2, \ \ell_L/a_s\gg \sqrt{v_L/v_s}$
\begin{equation}\label{dissipation}
\frac 1 {6\pi\mu} \int P dt\approx \frac{2
a_s\ell_L^2}{T_\ell}\left(1+O\left(\e^2 \frac {v_L}{v_s} \,\frac
{T_\ell}{T_v}\right)\right),\quad T_\ell+T_v=\tau/2\,
\end{equation}
where $T_\ell$ ($T_v$) is the time for traversing the horizontal
(vertical) leg. (Here $\e^2$ is actually $(a_s/\ell_L)^2$ rather
then the much larger $(a_L/\ell_s)^2$. Also note that the second
term in Eq.~(\ref{metric}) contributed $O(v_L/T_\ell)$ rather then
$O(v_L^2/(v_sT_\ell))$ as one may have expected from
Eq.~(\ref{metric}) which is dominated by the small volume.) The
optimal strategy, in this range of parameters, is to spend most of
the stroke's time on extending $\ell$. By
Eqs.~(\ref{delta},\ref{step},\ref{dissipation}) this gives the
drag
\begin{equation}\label{delta-best}
\delta \approx 4 a_s
\end{equation}  where $a_s$ is the
radius of the small bladder. {\em This allows for the transport of
a large sphere with the drag determined by the small sphere.} To
beat dragging, we need $a_s=a/4$, which means that most of the
volume, $63/64$, must be shuttled between the two bladders in each
stroke.

It is instructive to compare Pushmepullyou with the swimming
efficiency of models of (spherical) micro-organisms that swim by
beating flagella. These have been extensively studied by the
school of Lighthill and Taylor \cite{lighthill,blake} where one
finds $\delta \ge 100\, a$. This is much worse than dragging. (We
could not find estimates for the efficiency $\delta$ for swimming
by ciliary motion \cite{cilia}, but we expect that they are rather
poor, as for other squirmers \cite{agk}.) For models of bacteria
that swim by propagating longitudinal waves along their surfaces
Stone and Samuel \cite{samuel} established the (theoretical) lower
bound $\delta \ge \frac{4}{3} a$. (Actual models of squirmers do
much worse than the bound.) If the pushmepullyou swimmer is
allowed to make large strokes, it can beat the efficiency of all
of the above.

Eqs.~(\ref{metric},\ref{delta-best})do not strictly apply to metaboly
because the viscosity of the fluid inside the organism can not be
neglected and presumably dominates the dissipation. Euglena are
not as efficient as Pushmepullyou.


It is likely that some artificial micro-swimmers will be
constrained to make only small (relative) strokes. Small strokes
necessarily lead to large drag \cite{agk}, but it is still
interesting to see how large. Suppose $\delta\log
\ell\sim\delta\log v,\; a_1\sim a_2$. The dissipation in one stroke is
then\begin{equation}\label{dissipation-small-stroke} \frac {\int P
dt} {6\pi
\mu}=(\delta\ell)^2\left(\frac{a}{T_\ell}\right)
\left(1+O\left(\e^2\frac{T_\ell}{T_v}\right)\right)
\end{equation}
>From Eq.~(\ref{curvature}) and noting that $T_\ell=\frac 1 2
\tau$, one finds
\begin{equation}\label{delta-squirmer}
\delta\approx \frac{72}{(\delta\log v)^2}\ a\ .
\end{equation}

We shall now outline how the key results,
Eqs.~(\ref{swim},\ref{metric}), are derived. The flow around a
pair of spheres is a classical problem in fluid dynamics which has
been extensively studied \cite{JO,KK}. We could have borrowed from
the general results, e.g. in \cite {JO}, and adapt them to the
case at hand. However, it is both simpler and more instructive to
start from scratch: The classical Stokes solution \cite{batchelor}
describing the flow around a single sphere of radius $a$ dragged
by a force $f$ and, in addition, dilated at rate $\dot v$
\begin{equation}\label{stokes}\pi \vec{u}({\vec x};a,f,\dot v) = \frac 1
{6\mu|x|}\left(\left(3+\frac{a^2}{x^2}\right)\vec{f}+\left( 1
-\frac {a^2}{x^2}\right) 3(\vec{f}\cdot\hat x)\hat x\right)+ \frac
{\dot v}{ x^2} \hat x.\end{equation} $\vec{u}(\vec{x};a,f,\dot v)$
is the velocity field at a position $\vec{x}$ from the center of
the sphere. The left term is the known Stokes solution. (A
Stokeslet, \cite{batchelor}, is defined as the Stokes solution for
$a=0$.) The term on the right is a source term.

Since Stokes equations are linear, a superposition of the
solutions for two dilating spheres is a solution of the
differential equations. However, it does not quite satisfy the
no-slip boundary condition on the two spheres: There is an error
of order $\e $. The superposition is therefore an approximate
solution provided the two spheres are far apart.

The (approximate)  solution determine the velocities $U_i$ of the
centers of the two spheres:
\begin{equation}\label{U}
U_i= \vec{u}(a_i
\hat{f};a_i,(-)^jf,0)+\,\vec{u}((-)^i\ell\hat{f};a_j,(-)^if,(-)^i\dot
v),\quad i\neq j\in\{1,2\}
\end{equation}
The first term on the right describes how each sphere moves
relative to the fluid according to Stokes law as a result of the
force $\vec f$ acting on it. The second term (which is typically
smaller) describes the velocity of the fluid surrounding the
sphere (at distances $\gg a$ but $\ll\ell$) as a result of the
movement of the other sphere. By symmetry,  the net velocities of
the two sphere and the net forces on them are parallel to the axis
connecting the centers of the two spheres, and can be taken as
scalars. To leading order in $\e$ Eq.~(\ref{U}) reduces to
\begin{equation}2\pi U_i=(-)^j \frac f \mu \left(\frac 1 {3 a_i}-\frac 1 {2
\ell}\right) +\frac {\dot v}{2 \ell^2}
\end{equation}
Using $ \dot \ell =-U_1+U_2$ gives the force in the rod
\begin{equation}\label{force}
  f =-6\pi\mu \left(\frac 1 {a_1} +\frac 1{a_2}\right)^{-1} \  \dot \ell
\end{equation}
Dropping sub-leading terms in $\e $ gives Eq.~(\ref{swim}).

We now turn to Eq.~(\ref{metric}). Consider first the case $\dot
v=0$. The power supplied by the rod is $-f(U_2-U_1)=-f\dot \ell$
which  gives the first term. Now consider the case $\dot\ell=0$.
The stress on the surface of the expanding sphere is given by
\begin{equation}\label{stress}
\sigma=-\frac{2\mu \dot v}{4\pi}\, \left(\frac 1
{x^2}\right)^\prime=\frac{\mu\dot v}{\pi a^3}
\end{equation}
The power requisite to expand one sphere is then
\begin{equation}\label{dilating}
4\pi a^2\sigma\dot a=\sigma \dot v =\frac{4\mu}{3 v} (\dot v)^2
\end{equation}
Since there are two spheres, this give the second term in
Eq.~(\ref{metric}).

There are no mixed terms in the dissipation proportional to
$\dot\ell\dot v$. This can be seen from the following argument. To
the leading order in $\e^0$, which is all we care about, the
metric must be independent of $\ell$, (see Eq.~(\ref{metric}).
Sending $\ell\to -\ell $ is equivalent to exchanging the two
spheres. This can not affect the dissipation and hence the metric
must be even function of $\dot \ell$.  In particular, there can
not be a term $\dot v \dot\ell$ in the metric. This completes the
proof of Eq.~(\ref{metric}).

{\bf Acknowledgment} This work is supported in part by the EU
grant HPRN-CT-2002-00277. We thank H. Berg, H. Stone, and
especially Richard Triemer
for useful correspondence and for the Euglena racing tests.

\end{document}